\documentclass[letter,11pt]{article}
\usepackage[utf8]{inputenc}

\usepackage{amsmath,amssymb,amsfonts}
\usepackage{graphicx,xcolor}
\usepackage{xurl}

\usepackage[authoryear]{natbib}
\usepackage[parfill]{parskip} % Begin paragraphs with an empty line rather than an indent

\usepackage{booktabs} % for much better looking tables
\usepackage{array} % for better arrays (eg matrices) in maths
\usepackage{paralist} % very flexible & customisable lists (eg. enumerate/itemize, etc.)
\usepackage{verbatim} % adds environment for commenting out blocks of text & for better verbatim
\usepackage{subfig} % make it possible to include more than one captioned figure/table in a single float

\usepackage[nottoc,notlof,notlot]{tocbibind} % Put the bibliography in the ToC
\usepackage[titles,subfigure]{tocloft} % Alter the style of the Table of Contents

\title{A Case for AXI}
\author{Viktor Tsyrennikov\\Quantitative Research and Analysis\footnote{E-mail: viktor.tsyrennikov@quantific.us. The author thanks Joel Shapiro, Sven Klinger, Marcus Burnett, Vita Faychuk, and several anonymous reviewers.\newline The author gratefully acknowledges financial and data support from SOFR Academy, Inc.}}
\date{June 30, 2025}

\begin{document}
\maketitle

\begin{abstract}
In the LIBOR era, banks routinely tied revolving credit facilities to credit-sensitive benchmarks. This study assesses the Across-the-Curve Credit Spread Index (``AXI'')—a transparent, transaction-based measure of wholesale bank funding costs—as a complement to SOFR, summarizing its behavior, construction, and loan-pricing implications. AXI aggregates observable unsecured funding transactions across short- and long-term maturities to produce a daily credit spread that is IOSCO-aligned and operationally compatible with SOFR-based infrastructure.

The Financial Conditions Credit Spread Index (``FXI'') is a broader market companion to AXI and serves as its fallback. %FXI co-moves closely with AXI in normal times; during stress episodes affecting both the banking sector and the broader economy, their correlation exceeds 0.9.
FXI co-moves closely with AXI in normal times; under stress, the correlation of daily changes exceeds 0.9 for economy-wide shocks and remains strong in bank-specific stress, around 0.8 during the Silicon Valley Bank episode.

Empirically, AXI is strongly correlated with standard credit-spread measures and market-stress indicators, and is inversely related to financial-sector performance. SOFR\,+\,AXI exhibits correlations with macroeconomic variables with the signs and magnitudes expected of a credit-sensitive rate. In loan-pricing applications, SOFR\,+\,AXI reduces funding risk and can support spread discounts of up to 65 basis points without lowering risk-adjusted returns. In stress scenarios, banks relying on SOFR-only pricing can fail to recover as much as 15 basis points on revolving credit lines over as little as three months.

Taken together, AXI restores the credit sensitivity lost in the USD LIBOR transition while avoiding reliance on thin short-term markets, delivering significant economic value.

\end{abstract}

\newpage
\tableofcontents
\newpage

\section{Introduction} %: The Case for a Credit-Sensitive Benchmark}
For decades, London InterBank Offered Rate (``LIBOR'') was the de facto benchmark for pricing a vast array of loans, derivatives, and bonds, peaking at~\$400 trillion in contracts globally.\footnote{This paper considers only U.S. dollar LIBOR.} LIBOR was an interbank credit-sensitive rate -- it embedded the credit risk of banks borrowing from each other without collateral. This feature historically helped banks and lenders manage funding risk: as a bank’s own funding costs rose (with its credit risk), LIBOR tended to rise in tandem. A floating loan rate tied to LIBOR would increase when the lender’s funding became more expensive, thereby mitigating net interest margin compression and reducing borrowers' incentives to draw on their credit lines during stress. Thus, reference rates with a bank credit risk component can hedge liquidity/funding risk in loan contracts. However, a manipulation scandal combined with low transaction volume undermined LIBOR's integrity and prompted a global transition to alternative rates.

In the U.S., the Alternative Reference Rates Committee (``ARRC'') selected the Secured Overnight Financing Rate (``SOFR''), a nearly risk-free overnight repurchase agreement (``repo'') rate, as LIBOR’s successor. SOFR is robust and transaction-based, but it lacks the bank credit risk element inherent in LIBOR. The Bank for International Settlements (``BIS'') noted that risk-free rates (``RFRs''), while credible, “do so at the expense of not capturing banks’ marginal term funding costs,” suggesting that banks may need credit-sensitive benchmarks alongside a risk-free rate.\footnote{Here, ``marginal'' indicates that AXI measures the incremental funding cost over the risk-free rate.}

The trio of U.S. financial regulators Federal Reserve–FDIC–OCC (2020) explicitly did not endorse a single replacement and said banks may choose rates “most appropriate for their circumstances” so long as they are robust. This opened the door to credit-sensitive rates (``CSR'') like Bloomberg’s BSBY, Ameribor, and Across-the-Curve Credit Spread Index (``AXI''), even as regulators warned any new benchmark must meet IOSCO principles to avoid another LIBOR-like fiasco.

U.S. banks now rely mostly on deposits and longer-term debt, rather than the short-term wholesale funding that LIBOR reflects. Indeed, since the 2008 crisis, U.S. banks’ share of funding from 1-year unsecured debt dropped to roughly 2–3\%, suggesting that banks can operate with a risk-free base rate in normal conditions; see \citet{BowmanScottiVojtech2020LIBORFunding}.

In stress, however, funding costs spike. The LIBOR-OIS spread increased 360 bps in 2008 and 140 bps in 2020. This indicates the value of a credit-sensitive uplift in those rare but critical moments. In 2019, a group of regional banks warned that a ``SOFR-only environment'' could exacerbate drawdowns under stress, forcing banks to curtail lending or charge higher upfront spreads.\footnote{See \citet{MarshallEtAl2019CreditSensitivity} for the open letter by ten large regional banks to the Federal Reserve's Vice‑Chair, Comptroller of the Currency, and FDIC Chair.} \citet{Cooperman2022} argued that if loan rates do not rise when funding access tightens, borrowers have greater incentive to draw heavily on credit lines, implying larger drawdowns in a crisis under SOFR-only pricing.\footnote{\citet{Ghamami2023} argued that lack of credit sensitive benchmarks may reduce credit markets efficiency and disproportionately impact small and medium enterprises that are more reliant on credit lines.}

Banks anticipating this would charge higher fixed spreads or reduce credit supply. In their model, it is optimal to use a reference rate with 70–80\% of LIBOR's credit sensitivity.\footnote{Perfect credit sensitivity is typically not optimal as credit demand is lower then. The model also allows for some funds to be redeposited with the originating bank, reducing its net liquidity outflow.} The demand impact of incorporating credit sensitivity appears limited. \citet{Jermann2024} estimates an ex-ante willingness to pay of 16.8 basis points for LIBOR-indexed borrowers to switch to SOFR indexing. This supports the use of SOFR plus a credit spread capturing the benefit of LIBOR's risk sensitivity without relying solely on volatile short-term markets.

Both U.S. and international stakeholders recognized the need for robust benchmarks post-LIBOR. Pure RFRs like SOFR are preferred for integrity, but credit-sensitive complements are useful for products where lender funding costs and systemic risk are relevant. This sets the stage for developing indices such as the Across-the-Curve Credit Spread Index (``AXI'') that attempt to combine robustness with credit sensitivity.

The following section explains the key benefits of AXI and its methodology. Section~\ref{sec-axi-drivers} analyze its empirical properties via correlations with major credit indexes, core macroeconomic variables, and measures of economic uncertainty. To quantify AXI benefits, Section~\ref{sec-axi-value} analyzes profitability of hypothetical loans during the recent stress periods and describes the loan rate discount that banks using AXI could offer due to reduced funding risk.

\section{Design of AXI}
AXI is a composite credit spread intended, as a credit-sensitive ``add-on” to SOFR for lending products, flexible rate notes, and derivatives. It was developed by \citet{Berndt2023}. SOFR Academy implemented the index and positions it as a potential market standard for U.S. commercial and industrial (``C\&I'') loans, credit lines, and related derivatives.

\begin{figure}[!htb]
	\includegraphics[]{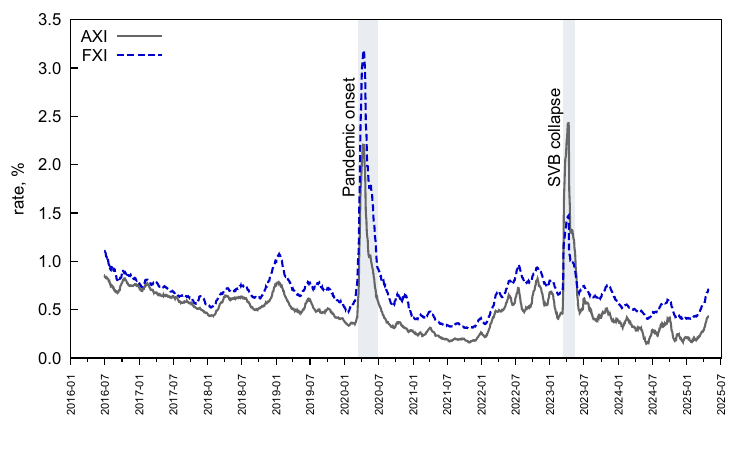}
	\caption{AXI and FXI}
	\label{fig-rates-fxi}
\end{figure}

AXI measures the recent average cost of wholesale unsecured debt funding for U.S. banks, across the maturity curve: from overnight to 5 years. It is the yield spread banks must pay over the risk-free rate to raise funds ``across the curve.'' Using maturities beyond one year contrasts with LIBOR and other credit sensitive proposals,\footnote{\citet{IOSCO2023Libor} reviewed the existing alternatives, including BSBY and Ameribor, and concluded that they should ``refrain from any representation that [they] are `IOSCO-compliant'.'' BSBY was discontinued in November 2024.} which are based solely on short-term borrowing -- 12 months or less. Additionally, unlike LIBOR that polled a small panel of banks about hypothetical short-term borrowing rates, AXI is based on a large number of actual trades.

AXI is calculated as a volume-weighted average of credit spreads on various bank debt instruments: certificates of deposit (CDs), commercial paper, wholesale term deposits, interbank loans, and -- for longer maturities -- corporate bonds and notes of banks. So, the index dynamically ``adapts'' if banks shift their funding mix. Figure~\ref{fig-rates-fxi} shows the full history of AXI.

%The weights reflect both issuance volumes and transaction volumes in each maturity segment, so the index adapts if banks dynamically shift their funding mix. Notably, to ensure transparency, AXI calculation is based on public market transactions: TRACE for bond trades and CP/CD trading data for commercial paper.

This design has several advantages:
\begin{itemize}
	\item[A.] Accuracy and Transparency\newline
	Currently, large banks rely more on issuing term bonds and notes than on overnight interbank loans. Since the 2008 crisis, banks ``no longer fund themselves at LIBOR,'' as they term out funding to reduce rollover risk.\footnote{According to \citet{Schrimpf2019}, a study from the BIS ``banks have lengthened their funding maturities, incentivised by the Net Stable Funding Ratio.''} AXI also captures the longer maturity funding, so it should track banks’ true marginal funding costs better than LIBOR.

	The calculation is based on public market transactions: Trade Reporting and Compliance Engine for bond trades and CP/CD trading data for commercial paper from Depository Trust and Clearing Corporation.\footnote{In addition to including longer maturities, AXI's transaction base covers all banks. In contrast, the discontinued BSBY's transaction base covered mostly by large banks and Ameribor's covers mainly regional and community banks. Further, Ameribor relies only on transactions on the American Financial Exchange, which are not included in the determination of AXI.}

	\item[B.] Stability and Robustness\newline
	By averaging across the curve, AXI smooths out idiosyncratic swings that may affect only one segment of the market. Thus, impact on AXI of a temporary liquidity squeeze in the one-month CP/CD market would be limited if longer-term funding markets remained steady. This reduces the likelihood of sharp transient spikes that ``plagued LIBOR'' during liquidity events. It is constructed to be IOSCO-compliant\footnote{\citet{Promontory2024AXIFXIReview} audit found AXI and FXI adheres to key IOSCO principles for benchmark design (covers AXI and FXI methodology), data sufficiency, and transparency of benchmark determination.} and to remain available in all market conditions.

	\item[C.] Flexibility\newline
	AXI is not a standalone interest rate, but a spread that can be added to a SOFR average or term rate to create a credit-sensitive benchmark yield. It is a transaction-based analog to the LIBOR-OIS spread. Being a spread, AXI offers the full spectrum of credit sensitivity, not allowed by LIBOR, as banks can add a portion of AXI to loan rate:
	\begin{equation}
		\textit{credit sensitive rate} = \textit{reference rate} + \textit{fixed spread} + c \cdot \textit{AXI}.
		\label{eq-sofrx}
	\end{equation}

	Credit sensitivity parameter $ c $ is bounded between 0 and 1. The resulting rate is a hybrid between a fully credit sensitive benchmark like LIBOR and a risk-free rate like SOFR.

\end{itemize}

AXI represents banks’ marginal cost of funds by construction. It adds a credit-sensitive layer for banks that need it, e.g., the U.S. regional and mid-sized banks, to align loan yields with their own funding risk. The largest banks (GSIBs) with ample deposits may be content with SOFR-only pricing, as suggested in \citet{Cooperman2025}, or use lower credit sensitivity parameter.

\subsection{FXI, a broad market companion to AXI}
The Financial Conditions Credit Spread Index (``FXI'') measures the marginal cost of unsecured wholesale debt funding faced by U.S. institutions. It includes all wholesale funding transactions that meet the same size, tenor, and data-quality filters used for AXI. Accordingly, relative to the bank-specific AXI, FXI provides a broader, macroeconomic gauge of funding conditions. FXI generally exceeds AXI except during periods of bank-specific stress, as Figure~\ref{fig-rates-fxi} illustrates.

Both indices capture stress in their respective markets, and divergence between them indicates whether turmoil is confined to banks or spilling over to the broader economy.

Given its much larger transaction pool, FXI is more statistically robust than AXI and provides a stronger basis for derivative markets to build on. By construction, AXI and FXI share a common methodology, making FXI a natural fallback reference for AXI.\footnote{AXI and FXI methodologies are available at, respectively, \url{https://www.invescosofracademyaxi.com/dam/jcr:2574214c-341c-4813-855a-b2f6adc417e0/Invesco-SOFR-Academy-USD-Across-the-Curve-Credit-Spread-Index-May-2025.pdf} and \url{https://www.invescosofracademyaxi.com/dam/jcr:a442351c-5809-4989-bd2f-3c1b4e4d6c66/Invesco-SOFR-Academy-USD-Financial-Conditions-Credit-Spread-Index-May-2025.pdf}.} A sufficiently severe bank-funding shock is likely to propagate to the broader economy, pushing FXI higher alongside AXI. Consistent with this, daily changes in FXI and AXI are strongly positively correlated:
\begin{equation*}
    cor(\Delta\textit{AXI},\ \Delta\textit{FXI}) = 0.777.
\end{equation*}
This correlation increases during economy-wide stress: during the onset of the COVID-19 pandemic (March 1–June 30, 2020) it was 93.6\%, whereas during the Silicon Valley Bank (SVB) episode (March 1–June 30, 2023) it was 80.0\%, close to the full-sample estimate.

\subsection{A brief on the index methodology}
Each transaction underlying the index is characterized by rate, maturity, and dollar size (volume). Transactions are aggregated into the short-term (maturity 0-1 year) and long-term (maturity 1-5 years) components.

For a given maturity bucket $ m $, the dollar-volume-weighted median spread is computed. Short-term (ST) spread corresponds to the bucket 0-1 year. Long-term (LT) spread is the volume-weighted average of the median spreads for buckets 1-2 years, 2-3 years, 3-4 years, and 4-5 years.%\footnote{This construction is numerically close to the volume-weighted median for maturities 1-5 year.}

Maturity-weighed dollar volume is used as weights. Letting $ j $ index long-term maturity buckets, the weights are computed as follows:
\small
\begin{align*}
	\text{ST weight}_t &=
	\frac{\text{ST maturity}_{t} \cdot \text{ST volume}_{t}}{\text{ST maturity}_{t} \cdot \text{ST volume}_{t} + \sum_j \text{LT maturity}_{jt} \cdot \text{LT volume}_{jt}},\\
	\medskip
	\text{LT weight}_{jt} &=
	\frac{\text{LT maturity}_{jt} \cdot \text{LT volume}_{jt}}{\text{ST maturity}_{t} \cdot \text{ST volume}_{t} + \sum_j \text{LT maturity}_{jt} \cdot \text{LT volume}_{jt}}.
\end{align*}
\normalsize
Maturity weighting recognizes that longer-term transaction reflect expected economic conditions rather than short-term liquidity concerns. Intuitively, a $T$-period loan could be viewed as a sequence of $T$ one-period loans of equal notional.

\begin{figure}[!htb]
	\includegraphics[]{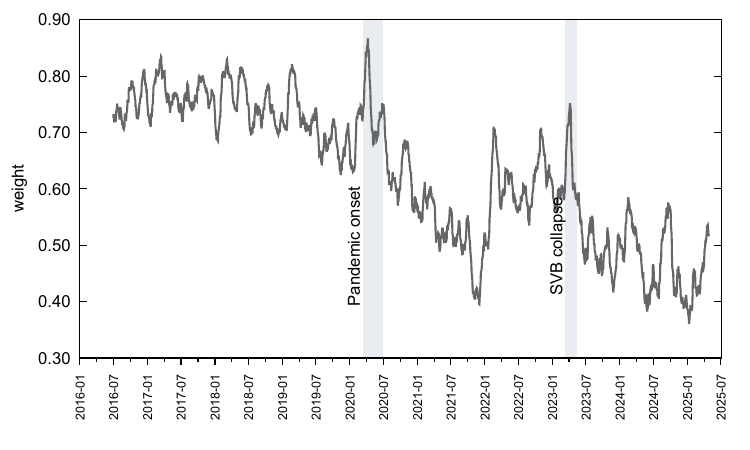}
	\caption{AXI weight on long-term (LT) transactions}
	\label{fig-axi-lt_fraction}
\end{figure}

Figure \ref{fig-axi-lt_fraction} shows the combined weight for long-term transactions.\footnote{With access only to the total LT volume and maturity, the share of LT transactions was computed as $ \text{LT weight}_t = \frac{1}{21}\sum_{k=1}^{21} \frac{\text{LT maturity}_{t-k} \cdot \text{LT volume}_{t-k}}{\text{ST maturity}_{t-k} \cdot \text{ST volume}_{t-k} + \text{LT maturity}_{t-k} \cdot \text{LT volume}_{t-k}} $.} The share of long-term transactions is always significant and above 36.0\%.\footnote{By design, the high LT weight is achieved through 21-business day averaging. Without the averaging, the LT weight is still significant but experiences occasional single-day declines driven by month-end balance sheet management activities and industry events, such as the transition to a T+1 settlement cycle on May 28, 2024.}

Long-term and short-term maturity-weighed daily dollar volumes are currently approximately balanced, see figure \ref{fig-axi-lt_fraction}. The correlation between the two maturity-weighted volumes is insignificant at 0.076, similar to the non-weighed correlation of 0.069. It suggests that the long-term and short-term markets are independent: a buyer in one segment is neither more nor less likely to be a buyer or seller in the other segment.

The weighted average spread based on a single day transactions is:
\begin{equation}
	\text{Spread}_t = \sum_j \text{LT spread}_{jt} \cdot \text{LT weight}_{jt}  + \text{ST spread}_t \cdot \text{ST weight}_t .
	\label{eqn-axi-definition}
\end{equation}
To smooth the daily variation and achieve index robustness, daily spreads are averaged over the previous 21 business days:
\begin{equation}
	\text{AXI}_t = \frac{1}{21} \sum_{j=1}^{21} \text{Spread}_{t-j}.
	\label{eqn-axi-definition}
\end{equation}
In practice 21-business day averaging is numerically similar to a 30-calendar day averaging used by SOFR.\footnote{The Alternative Reference Rates Committee (``ARRC'') has explicitly highlighted compounded average as the preferred SOFR formulation for most applications, noting that ``30-day average of SOFR would be sufficient to smooth most of the day-to-day volatility,'' see \citet{ARRC2019SOFRARM}.}

%Thus, this article focuses on the following index:
%\begin{equation}
%	\textit{SOFRx 30-day} = \textit{SOFR 30-day compounded average}\ + \textit{ AXI spread}.
%	\label{eq-sofrx}
%\end{equation}

\subsection{Trading volume underlying AXI}
The trading volume underlying AXI has been substantial throughout the history.
Between June 2016 and March 2025, the total dollar volume averaged at \$463.1 billion with the lowest ``non-holiday'' volume of \$265.4 billion observed on November 10, 2024. Volume fluctuations are driven by economic news. For example, the largest volume of \$684.5 billion was recorded on June 14, 2021, coinciding with the reopening of the pandemic-affected U.S. economy. Importantly, there is a structural break in H1 2021 when banks experienced an unprecedented inflow of unstable ``hot money'' deposits, see \citet{FederalReserve2021}. Not meeting the leverage ratio and net stable funding ratio regulatory constraints, banks turned to the short-term whole-sale funding markets. Long-term transaction volume peaked in 2021 and has been declining gradually since, likely due to heightened economic uncertainty brought by the Covid-19 pandemic and subsequent inflation spike. Importantly, the increase in short-term dollar volume was nearly offset by the decrease in maturity, and the fraction of long-term transaction remained relatively stable in Figure ~\ref{fig-axi-lt_fraction}.

% May 28, 2021 - 6T budget by Biden

% Dollar transaction volume increased significantly during the first half of 2021. During this time the Federal Reserve increased the use of repurchase agreements to continue bolster liquidity in the economy battered by the pandemic, while the economy started to recover and demand for short-term funding increased. At the same time, SOFR was set to replace LIBOR in all new contracts at the end of 2021.

\begin{figure}[!htb]
	\includegraphics[]{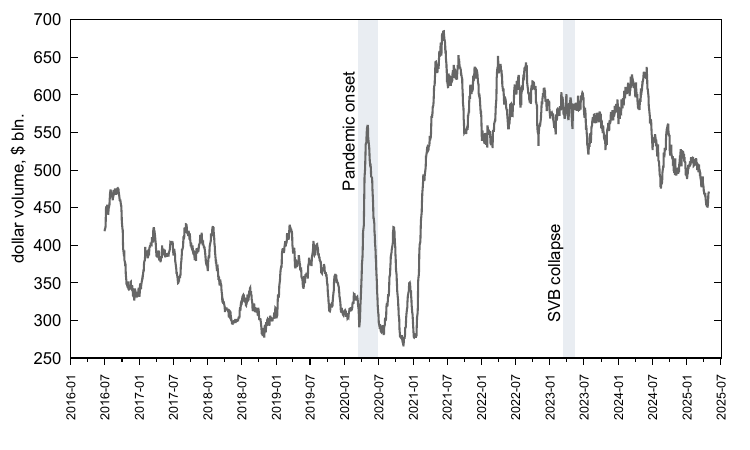}
	\caption{Dollar volume of transactions underlying AXI}
	\label{fig-axivolume}
\end{figure}

The unweighted (``raw'') volatility of the AXI underlying dollar volume is 1.2\%. It is lower than volatility of SOFR dollar volume of 4.2\% because AXI is based on transactions from the past 21-business days and SOFR on a single business day. The low volatility of raw trading volume means there is a stable base of transactions to support AXI.

\begin{figure}[!htb]
	\includegraphics[]{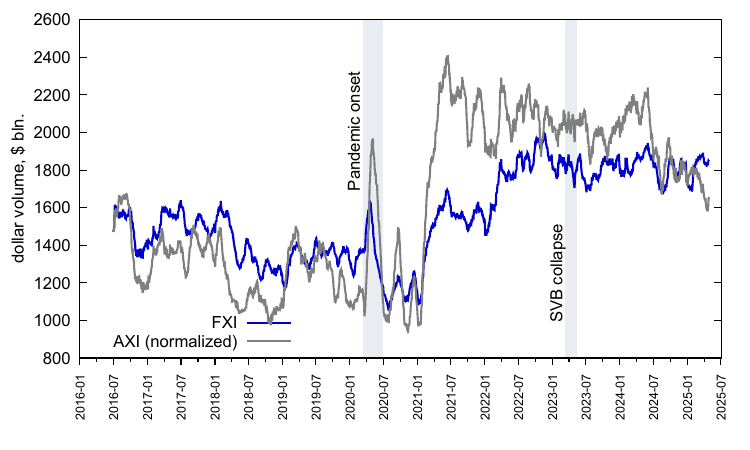}
	\caption{Dollar volume of transactions underlying FXI}
	\label{fig-fxi-volume}
\end{figure}

FXI, that also serves as a fallback for AXI, is supported by much higher volume: \$1,558 billion on average and greater than \$1 trillion at all times. Thus, high volatility of the AXI underlying daily dollar volume is not a concern. Further, as Figure~\ref{fig-fxi-volume} shows, dollar volumes of AXI and FXI exhibit similar dynamics except for the structural break in H1 2021.\footnote{AXI was normalized to begin at the same value as FXI.}

Incorporating long-term spreads is one of AXI and FXI's key benefits. For both indexes, the long-term spread is closer to the index value. It is also significantly more stable as measured by the coefficient of variation, see Table~\ref{tbl-lt_st_spreads}.

\begin{table}[!htb]
	\centering

	A. AXI spreads
	\medskip

	\small
	\begin{tabular}{l|cc|cc|c}
%		statistic & AXI & Daily spread & LT spread & ST spread & LT weight \\
		          &     & Daily  & LT     & ST     & LT\\
		statistic & AXI & spread & spread & spread & weight \\
		\hline
		mean          & 0.5165 & 0.5168 & 0.7714 & 0.0723 & 0.631 \\
		st. deviation & 0.2999 & 0.3861 & 0.4189 & 0.2052 \\
		coef. of variation & 0.5806 & 0.7471 & 0.5430 & 2.8375 \\
	\end{tabular}
	\normalsize

	\bigskip
	B. FXI spreads
	\medskip

	\small
	\begin{tabular}{l|cc|cc|c}
%		statistic & FXI & Daily spread & LT spread & ST spread & LT weight \\
		          &     & Daily  & LT     & ST     & LT\\
		statistic & FXI & spread & spread & spread & weight \\
		\hline
		mean               & 0.6905 & 0.6923 & 0.8280 & 0.1203 & 0.805 \\
		st.deviation       & 0.3200 & 0.3784 & 0.4104 & 0.2209 \\
		coef. of variation & 0.4635 & 0.5465 & 0.4957 & 1.8363 \\
	\end{tabular}
	\caption{Statistical properties of short- and long-term spreads}
	\label{tbl-lt_st_spreads}
\end{table}

%The short-term spread is lower than the long term spread, i.e., the term curve is upward sloping as expected.\footnote{The curve did invert during the federal funds liquidity event in September 2019 and 4 days in April 2023. On April 23, 2023, when the short-term spread exceeded long-term spread by 0.77\%, the U.S. economic growth was reported to slow down significantly.}

%Volatility of the maturity-weighed dollar volumes growth, measured as daily percentage changes, are 57.7\% and 55.8\%, respectively for the short-term and long-term segments. They are comparable and also economically significant. The unweighted volatility of the AXI underlying dollar volume is 35.9\%. For comparison, volatility of dollar trading volume underlying the Overnight Bank Funding Rate (``OBFR''), measuring the cost of unsecured short-term bank borrowing in the U.S., is significantly lower at 15.7\%. High volatility of the volume is not a concern given that AXI falls back to FXI, and the latter's underlying volume averaged at \$74 billion with un-weighted volatility of 22.2\%.

% The largest short-term dollar volume peaked on September 22, 2022, i.e., around the time of a major rate hike by the U.S. Federal Reserve. The most active day for the long-term segment was February 16, 2017, a liquidity event affecting Treasury and repo markets. LT activity appears more sensitive to trade size and possibly episodic institutional flows.

\subsection{Credit-sensitive rates}

%AXI captures banks’ marginal cost of unsecured wholesale funding across the curve—from overnight out to five years.

When added to the risk-free SOFR, AXI yields a credit-sensitive rate for use in financial contracts. It is an unsecured-credit benchmark, as LIBOR was.\footnote{\citet{Rajiv2024} discussed various ways market participants can use the credit-spread reference benchmarks AXI and FXI.} This section compares SOFR+AXI with LIBOR, which is the relevant benchmark for two reasons. First, AXI was designed to aid LIBOR replacement. Second, even in 2022 -- after most new deals had migrated to SOFR -- about 60 percent of outstanding U.S. C\&I loans still referenced LIBOR, see \citet{Cooperman2025}.\footnote{After June 30, 2023, LIBOR series are proxied by term SOFR plus a fixed spread: the synthetic 3-month LIBOR is computed as 3-month term SOFR plus 11.48 basis points.}

Figure \ref{fig-axivolume} plots the nearly 10-year history from July 2016 until April 2025. Depending on the credit sensitivity parameter $ c $ in equation~\eqref{eq-sofrx}, SOFRx lies between SOFR (zero sensitivity: $ c=0 $) and SOFR+AXI (full sensitivity: $ c=1 $).\footnote{Overnight SOFR series starts on April 3, 2018. Federal Reserve Bank in New York extended the overnight SOFR series, and the extended SOFR 30-day compounded average was computed using the SOFR Methodology back to September 21, 2014.} The study uses 3-month Libor that was by far the most referenced rate according to \citet{ARRC2018Second} report.

While LIBOR series end on September 30, 2024, there were no new LIBOR contracts after December 31, 2021. Publication of key LIBOR tenors continued until June 30, 2023 when the U.K.'s Financial Conduct Authority declared ``loss of representatives.'' Thereafter, LIBOR was proxied by term SOFR plus fixed spread.

SOFR+AXI is always above LIBOR, and the average basis between the two is 0.37\%, and it is material.\footnote{If the data after December 21, 2021 is excluded, the average spread is nearly the same at 0.41\%.}
\begin{equation*}
	\textit{average}(\text{SOFR+AXI} - \text{LIBOR)} = 16\ \textit{basis points}.
\end{equation*}
With $ c=2/3 $ sensitivity, the average spread between SOFR+(2/3)AXI and LIBOR is -0.01\%. That is, banks would maintain the same expected profitability as under LIBOR if loans were indexed to the new credit sensitive rate.

Although SOFR+AXI and LIBOR moved broadly together over time, correlation between their daily changes is weak at 0.133.\footnote{With temporal aggregation, the correlation rises, reaching 0.878 at the quarterly frequency.} This indicates that the two benchmarks respond to economic risk factors differently, with the largest divergences occurring during stress episodes.

% AXI's average is only 0.52\% against SOFR's 2.62\%, and AXI's daily volatility is 6.5 times lower. Changes in AXI and SOFR 30-day compounded average are effectively uncorrelated at the the daily frequency.

During the Pandemic onset, LIBOR declined below SOFR while SOFR+AXI jumped, and their difference expanded above 2\%. During the SVB collapse, SOFR+AXI also jumped while LIBOR continued its trend, and their difference reached almost 2.5\%. LIBOR and SOFR are more strongly correlated than LIBOR and SOFR+AXI, further suggesting LIBOR has been less representative of bank borrowing costs. In both instances, linking loans to LIBOR could result in losses for banks contractually obligated to extend loans, even when rates fall below their funding costs.

\begin{figure}[!htb]
	\includegraphics[]{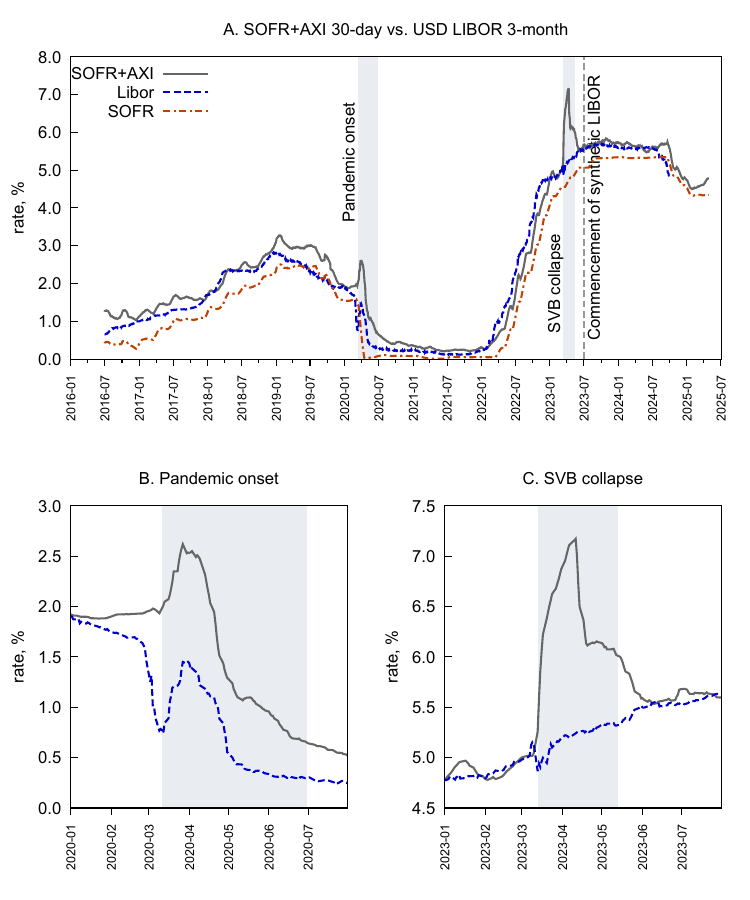}
	\caption{SOFR+AXI vs LIBOR}
	\label{fig-rates}
\end{figure}

\begin{figure}[!htb]
	\includegraphics{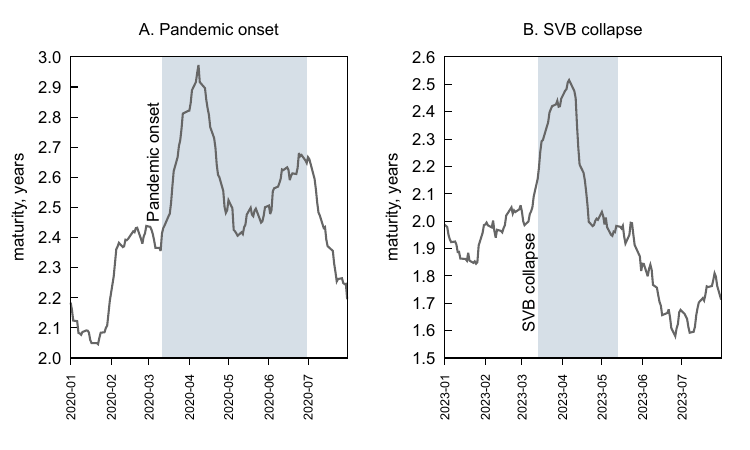}
	\caption{Weighted average maturity of transactions underlying AXI}
	\label{fig-rates-mat}
\end{figure}

Figure~\ref{fig-rates-mat} shows that the volume-weighted average maturity of AXI transactions lengthened in both stress episodes, driven primarily by higher volumes at longer maturities. Long-term dollar volume rose in both periods, as expected. By contrast, short-term volume dipped briefly at the onset of the pandemic and showed little response around the SVB collapse. Short-dated maturities also shortened during stress, as expected. Accordingly, AXI’s effective tenor lengthens when conditions tighten and normalizes as stress abates—a dynamic maturity adjustment that sets AXI apart from benchmarks with fixed tenor weights.

\subsubsection{SOFR+AXI correlations with core macroeconomic variables}
Table~\ref{tbl-correlations} reports correlations of SOFR+AXI and LIBOR with the Dodd-Frank Act Stress Testing (``DFAST'') 2025 core macroeconomic variables.\footnote{For variable definitions see \citet{FederalReserve2025DFASTDefinitions}.} Following \citet{Greene2017}, the table reports significance of the coefficient in a univariate linear regression. These correlations should be viewed only as ``suggestive,'' as this quarterly dataset is small covering the period from Q2 2016 to Q3 2024.\footnote{Both SOFR+AXI and LIBOR are transformed into quarterly differences, $ x_t-x_{t-1} $, as were other rates. Indexes HPI, CRE, DJIA, and VIX were log-differenced: $ ln(x_t)-ln(x_{t-1}) $. After differencing, there are 34 quarterly observations between Q3 2016 and Q4 2024 because AXI history is still short. For consistency, LIBOR series were restricted to the same period.} Variables with reported p-values below 0.05 are considered statistically significant, while p-value below 0.10 indicates marginal statistical significance.

With that in mind, the table shows the two benchmarks have largely the same drivers with one important difference. The 3-month Treasury bill rate and Prime rate exhibit the highest correlations, followed by BBB spread\footnote{Notice that SOFR+AXI rate is compared with a spread. It shows that SOFR+AXI indeed increases when corporate credit spreads measured by BBB spread increase.} and mortgage rate. LIBOR is also strongly correlated with the 5-year and 10-year Treasury note rates, while SOFR+AXI's correlations are lower and only marginally significant.

SOFR+AXI is negatively correlated with the inflation rate, i.e., it increases during economic stress typically associated with declining prices. Overall, it exhibits correlations of sign and magnitude expected of a credit-sensitive rate and similar to those of LIBOR. However, the experience during the recent stress periods suggests that LIBOR did not reflect banks' funding costs long before it has been discontinued.

\begin{table}[!htb]
	\centering
	\small
	\begin{tabular}{lrrrr}
		           &    \multicolumn{2}{c}{SOFR+AXI} & \multicolumn{2}{c}{LIBOR} \\
		  variable & correlation & p-value & correlation & p-value \\
		  \hline
                             Prime &   0.913 & \bf{0.000} &    0.924 & \bf{0.000} \\
                       Treasury 3m &   0.881 & \bf{0.000} &    0.967 & \bf{0.000} \\
                        BBB spread &   0.406 & \bf{0.017} &    0.600 & \bf{0.000} \\
                     Mortgage rate &   0.403 & \bf{0.018} &    0.680 & \bf{0.000} \\
                CPI inflation rate &  -0.363 & \bf{0.035} &   -0.168 &     0.341  \\
                       Treasury 5y &   0.316 &     0.068  &    0.675 & \bf{0.000} \\
                      Treasury 10y &   0.301 &     0.083  &    0.638 & \bf{0.000} \\
                 House price index &  -0.277 &     0.113  &   -0.194 &     0.272  \\
      Dow Jones Industrial Average &  -0.188 &     0.287  &   -0.314 &     0.070  \\
                Nominal GDP growth &  -0.160 &     0.364  &   -0.008 &     0.966  \\
                   Real GDP growth &  -0.121 &     0.497  &    0.027 &     0.881  \\
                 Unemployment rate &  -0.107 &     0.548  &   -0.269 &     0.125  \\
     Real disposable income growth &   0.076 &     0.668  &   -0.011 &     0.953  \\
                  Volatility index &   0.060 &     0.736  &   -0.009 &     0.959  \\
  Nominal disposable income growth &   0.056 &     0.754  &   -0.019 &     0.914  \\
                   CRE price index &   0.041 &     0.817  &    0.005 &     0.977  \\
	\end{tabular}
	\normalsize
	\caption{Correlations of quarterly changes in SOFR+AXI and LIBOR with the core macroeconomic variables}
	\label{tbl-correlations}
\end{table}

\section{Drivers of AXI}
\label{sec-axi-drivers}

AXI is a relatively small portion of SOFR+AXI and the correlations of SOFR+AXI with macroeconomic variables studied above are driven by SOFR to a significant extent. This section focuses on AXI's drivers, and Table~\ref{tbl-axi-drivers} lists potential candidates. All credit indexes are based on investment grade credit only, which is appropriate as banks' debt is considered high quality.\footnote{\citet{spglobal2024} report that financial companies have a significantly lower default rate than non-financial companies: 0.42\% vs. 1.88\% during 2014-2023. This rationalizes the focus on investment grade entities.} Like AXI, all credit indexes except for the credit market distress index were based on securities with maturity of 5 years or less.\footnote{The paper also analyzed Bloomberg FRN and US corporate indexes analogues based on all maturities, and the results were nearly unchanged.}

The analysis includes LIBOR-OIS spread (3-month tenor) that could be viewed as a LIBOR-age analog of AXI. The spread between 3-month LIBOR and T‑bill rate (``TED'') is the premium lenders demand to take bank credit risk. It was commonly used by large banks in internal capital planning and stress liquidity scenarios.

The analysis also includes several measures of economic uncertainty: the Federal Reserve Bank of New York's recession probability estimate, the economic policy uncertainty index, and volatility index. SPDR financial sector ETF was used as a proxy for the performance of the banking sector.\footnote{Using Standard and Poor's (``S\&P'') Financials Select Sector index (``S5FINL'') instead of XLF yielded nearly unchanged results.}

Table~\ref{tbl-correlations-fin} reports a contemporaneous correlation and correlations with lagged risk drivers. Again, reported p-values indicate a statistically significant relationship when below 0.05.\footnote{This analysis uses differenced AXI, CMDI, NFCI, EPU, TED, and logarithmically-differenced XLF and VIX. Changes in AXI and other indexes are then averaged weekly.}

\begin{table}[!htb]
	\centering
	\small
	\begin{tabular}{lcrr|rr}
		variable & stat. & current & lag 1 & lag 2 & lag 3 \\
		\hline
		Markit NA credit index
			& correl. & 0.028 & \color{blue}{\textbf{0.282}} & \textbf{0.288} & \textbf{0.275} \\
			& p-value & 0.551 & 0.000 & 0.000 & 0.000 \\
		Bloomberg FRN index OAS
			& correl. & \textbf{0.441} & \color{blue}{\textbf{0.451}} & \textbf{0.327} & \textbf{0.272}\\
			& p-value & 0.000 & 0.000 & 0.000 & 0.006\\
		Bloomberg FRN index
			& correl. & \textbf{-0.369} & \color{blue}{\textbf{-0.488}} & \textbf{-0.432} & \textbf{-0.299}\\
			& p-value & 0.000 & 0.000 & 0.000 & 0.000 \\
		Bloomberg US corporate index OAS
			& correl. & 0.399 & \color{blue}{\textbf{0.464}} & \textbf{0.430} & \textbf{0.336}\\
			& p-value & 0.000 & 0.000 & 0.000 & 0.000 \\
		Bloomberg US corporate index
			& correl. & -0.027 & \color{blue}{\textbf{-0.268}} & \textbf{-0.238} & \textbf{-0.233}\\
			& p-value & 0.562 & 0.000 & 0.000 & 0.000 \\
		National financial conditions index
		    & correl. & \textbf{0.458} & \color{blue}{\textbf{0.562}} & \textbf{0.562} & \textbf{0.453}\\
		    & p-value & 0.000 & 0.000 & 0.000 & 0.000 \\
		Credit market distress index
		    & correl. & \textbf{0.197} & \color{blue}{\textbf{0.245}} & \textbf{0.206} & \textbf{0.087} \\
		    & p-value & 0.000 & 0.000 & 0.000 & 0.064 \\
        LIBOR-OIS spread
		    & correl. & \textbf{0.461} & \color{blue}{\textbf{0.497}} & \textbf{0.359} & \textbf{0.179} \\
		    & p-value & 0.000 & 0.000 & 0.000 & 0.000 \\
        TED spread
		    & correl. & \textbf{0.397} & \color{blue}{\textbf{0.406}} & \textbf{0.345} & \textbf{0.269} \\
		    & p-value & 0.000 & 0.000 & 0.000 & 0.000 \\
		\hline
		SPDR financial sector ETF
			& correl. & \textbf{-0.107} & \color{blue}{\textbf{-0.292}} & \textbf{-0.256} & \textbf{-0.217}\\
			& p-value & 0.022 & 0.000 & 0.000 & 0.000 \\
		\hline
		Economic policy uncertainty
		    & correl. & 0.016 & \color{blue}{\textbf{0.121}} & \textbf{0.079} & 0.024 \\
		    & p-val. & 0.731 & 0.009 & 0.091 & 0.614 \\
		Volatility index
			& correl. & 0.005 & \color{blue}{\textbf{0.130}} & \textbf{0.131} & \textbf{0.197}\\
			& p-value & 0.909 & 0.005 & 0.005 & 0.000 \\
	\end{tabular}

	\caption{Correlations of AXI with selected financial indicators}
	\label{tbl-correlations-fin}
\end{table}

Markit’s North American Investment Grade Index (``CDX'') references a basket of credit-default-swap (CDS) contracts with maturity up to five years on 125 of the most liquid North American investment-grade entities, including major banks and insurance companies. Because both series reflect unsecured credit conditions, changes in CDX are positively correlated with changes in AXI (0.262).

Bloomberg’s U.S. Floating Rate Note Index (``BFRN'') tracks the total return of U.S. investment-grade floating-rate notes.\footnote{Notes issued before April 1, 2021 typically reference 3-month USD LIBOR; those issued on or after that date reference compounded SOFR.} Because BFRN is a total-return index rather than a spread measure, AXI is compared with the index’s option-adjusted spread (``OAS''), the constant spread that prices credit and option risk over an FRN’s life by equating modeled cash-flow present value to market price.\footnote{Formally, with time step $\Delta t$,
\begin{equation*}
    \textit{Market price}_j = \frac{1}{S}\sum_{s=1}^S \sum_{t=1}^T \textit{cash flow}_{j,s,t} \cdot \exp\left(-\sum_{\tau=1}^t (r_{s,\tau}+\textit{OAS}_j)\Delta t\right),
\end{equation*}
where $j,s,t$ index the security, simulation path, and time. Bloomberg reports index-level OAS as the market-value-weighted average of constituent OAS values.} Changes in OAS are strongly and positively correlated with changes in AXI (0.451 at one week lag). By contrast, changes in the total-return index are negatively correlated with changes in AXI (-0.488): when credit spreads widen (AXI rises), FRN discount rates increase and mark-to-market values fall, producing negative contemporaneous returns.\footnote{Netting out a compounded SOFR index from BFRN to form an excess-return proxy does not change the sign: spread-driven mark-to-market moves dominate the risk-free leg. See \citet{TuckmanSerrat2022} for FRN return attribution.}

Bloomberg’s U.S. Corporate Index is a total-return benchmark of investment-grade, fixed-rate U.S. dollar corporate bonds. As with BFRN, changes in its OAS co-move strongly with changes in AXI (0.464 when OAS lags AXI by one week).

LIBOR-era indicators, the LIBOR–OIS and TED spreads, also exhibit strong positive correlations with changes in AXI (0.497 and 0.406 at a one-week lag), as expected. These correlations are material but well below unity, reinforcing that LIBOR was an imperfect proxy for banks’ marginal funding costs.

Chicago Fed's National Financial Conditions Index (``NFCI''), introduced by \citet{BraveButters2011}, is a dynamic composite measuring how tight or loose overall U.S. financial conditions are. Its inputs, e.g., five-year financial CDS spreads, overlap with those used in AXI, so changes in AXI are strongly and positively correlated with changes in NFCI (0.562).

The Corporate Bond Market Distress Index (``CMDI'') focuses on investment-grade non-financial corporate-bond markets signaling impaired primary or secondary-market access for borrowers. Small changes in AXI do not necessarily translate into stress in the overall credit market. For this reason and because of excluding financial firms, its co-movement with AXI is weaker, though still positive (0.245).

Returns on the SPDR Financial Select Sector exchange traded fund (``XLF) serve as a sector-specific risk barometer and proxy of balance sheet health. Higher bank-funding costs and AXI typically weigh on financial-sector equity prices, giving rise to a negative correlation (-0.292).

Macroeconomic risk factors correlate with AXI only weakly. The Economic Policy Uncertainty (``EPU'') index of \citet{Baker2016EPU} and the S\&P 500 option-implied volatility index (``VIX'') reflect broad economic uncertainty.\footnote{EPU is based on the frequency with which major newspapers simultaneously reference the economy, policy, and uncertainty.} Heightened uncertainty is expected to widen premia that feed into AXI. But, they reflect the broad economy rather than bank-funding frictions, and their relationships with AXI are modest at 0.12-0.13.

Correlation analysis shows that AXI lags other indicators by one week, as the correlations in Table~\ref{tbl-correlations-fin} are highest when other indicators lag AXI one week. The association with NFCI and BFRN is strong for weekly series, suggesting AXI is an important credit indicator for the US economy.

Furthermore, these relationships are "predictively"-causal, as changes in the studied indicators (excluding EPU) Granger-cause changes in AXI at a 99\% confidence level.\footnote{As noted by \citet{Granger1969}, such predictive causality may exist in the presence of confounding factors and omitted variables, even without a true causal relationship.} The analysis also found no evidence of reverse causality, as changes in AXI do not Granger-cause changes in the drivers at either a 99\% or 95\% confidence level. This absence of reverse causality, combined with the predictive relationships identified, provides stronger statistical support for the causal direction.

%The relationships are likely ``predictively''-causal as changes in the studied indicators (except EPU) \citet{Granger1969}-cause changes in AXI at at 99\% confidence level.\footnote{Predictive causality may exist when there is no true causal relationship in the presence of confounding factors and omitted variables.} Additionally, there is no evidence of reverse causality. That is, changes in AXI do not Granger-cause at 99\%, or 95\%, confidence level. At the monthly frequency, neither the selected drivers lead AXi nor the reverse is true.

The monthly correlations in table \ref{tbl-correlations-fin-extended} in appendix \ref{s-correlations-fin-extended} are stronger but follow the same pattern as weekly correlations. Importantly, most of the correlations become contemporaneous, which suggests that the lag of AXI's effects is short. This could be explained by the banking sector's significant impact on the economy, which allows shocks to AXI to propagate rapidly. Correspondingly, no evidence was found for a leading or Granger-causal effect between the selected drivers and AXI at this monthly frequency.

\subsection{Stress periods}
The defining feature of SOFR+AXI is its behavior during stress periods. The most recent such periods are the onset of the Covid-19 pandemic (“Pandemic onset”) and the collapse of the Silicon Valley Bank (“SVB Collapse”).

In tranquil markets SOFR+AXI behaves like SOFR plus a stable spread. When funding markets are under stress, SOFR+AXI tracks banks’ true unsecured funding costs instead of falling away from them. This section discusses two recent stress episodes - the COVID-19 pandemic onset (``Pandemic Onset'', March 2020) and the Silicon Valley Bank failure (``SVB Collapse'', March 2023) -- to illustrate expected behavior of a credit-sensitive rate. The two stress periods are qualitatively different: the pandemic has an economy-wide impact, while the SVB collapse affected mostly the financial sector.

\subsubsection{Potential exposure}
Variable rate lending dominates banks' balance sheets. C\&I loans alone accounted for about \$2.8 trillion, or 15 percent of total bank credit in Q4 2024.\footnote{\citet{FederalReserveH8} reports the total bank credit of all commercial banks as of March 26, 2025 was \$18,154.6 billion. Of this amount, C\&I loans were \$2,779.5 billion, i.e., 15.3\%.} However, as \citet{Cooperman2022} note, the drawn portion of C\&I lines represented only 29\% of the \$1.9 trillion committed at end-2019 -- leaving banks on the hook to fund the remaining 71\% on demand.

\begin{figure}[!htb]
	\centering
	\includegraphics{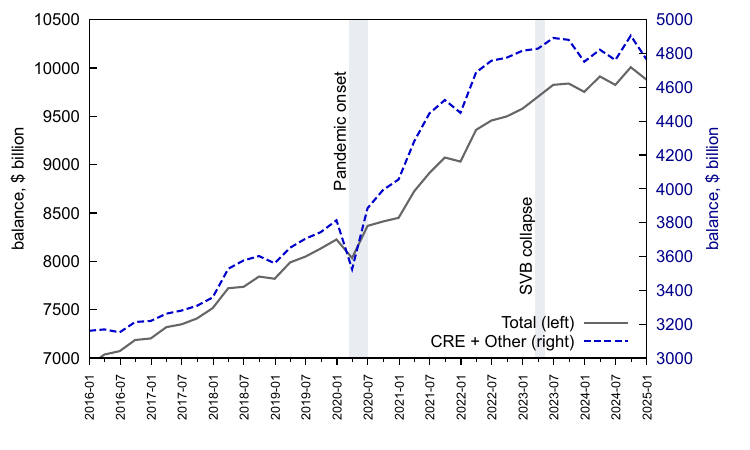}
	\caption{Unused commitments of banks}
	\label{fig-unused_commitments}
\end{figure}

Figure \ref{fig-unused_commitments} shows the scale of the potential exposure. Unused commitments of U.S. commercial banks, labeled ``Total'' in the figure, reached about \$10 trillion in 2024 Q4. The series labeled ``CRE + Other'' excludes commitments that are unlikely to be tapped under stress: credit card lines, home equity lines, and securities-underwriting facilities. By contrast, C\&I and CRE revolvers tend to be drawn aggressively in downturns, as seen in the 2008–09 crisis (\citet{IvashinaScharfstein2010,AcharyaSteffen2020}). Even after excluding the consumer and underwriting categories, the remaining unused lines still equal approximately 19.8\% percent of banking-sector assets, underscoring importance of using a credit-sensitive rate.

During the Pandemic Onset, banks saw their credit lines drawn down but not during the SVB collapse. At the same time, the exposure of banks as measured by the unused commitments -- as opposed to observed drawdowns -- during the two stress periods is substantial, with the majority accounted for by the largest 25 banks.

%Short-lived extreme stresses like the Pandemic onset and SVB collapse do not impact loans resetting monthly or less frequently, as majority of the impact on banks' cost of funds as measured by AXI disappears within a month.

\subsubsection{Pandemic onset}
This stress period began with the World Health Organization’s declaration of the start of a pandemic on March 11, 2020. It was characterized by a significant increase in uncertainty, as its potential toll on people and US economy were largely unpredictable. Starting on March 3, 2020, major media outlets started speculating on the impending move by the WHO, reflecting growing global concern about the virus’s spread.

During the Pandemic, AXI spike was driven by an increase in its long-term (“LT”) component that increased from about 0.4\% to 3.9\%. This change is economically significant and LT spread remained elevated until the end of June 2020.

%During this period, also the Treasury yield curve inverted, which is a strong predictor of a weakening economy cofounding with the effect of the Pandemic Onset. So, the duration of the Pandemic effect is likely shorter.

\begin{figure}[!htb]
	\centering
	\includegraphics{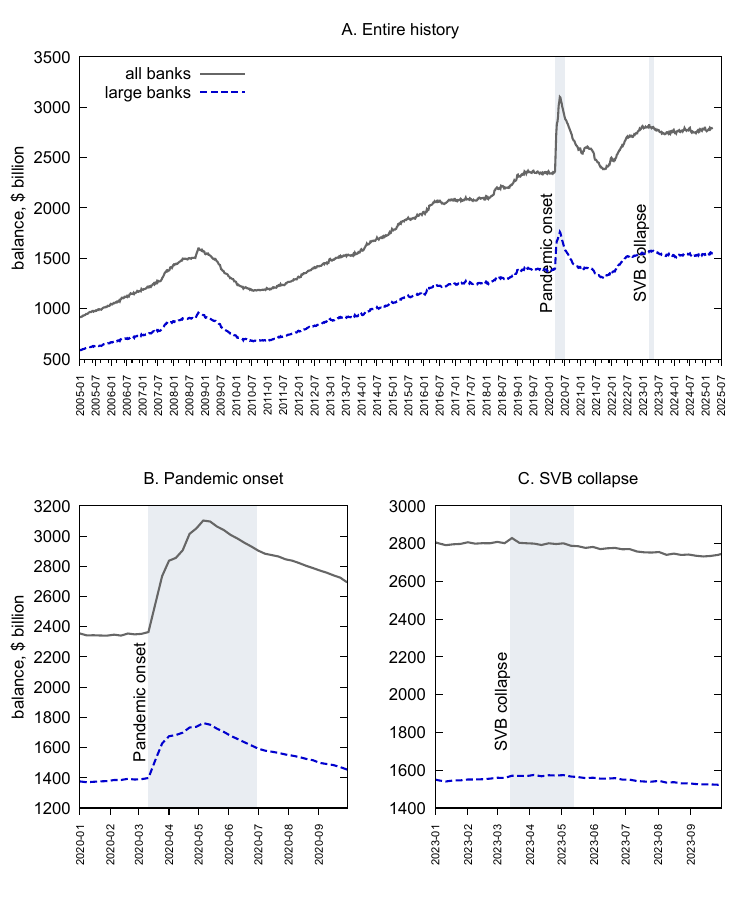}
	\caption{C\&I loan balance at the U.S. commercial banks}
	\label{fig-ci}
\end{figure}

Figure \ref{fig-ci} plots C\&I loan balances of all commercial banks in the US. They build up rapidly during stressful periods and follow the general economic trend otherwise. During the two weeks following the Pandemic onset, the C\&I balances increases by \$370 billion, and they peaked eight weeks from the Pandemic onset at \$738 billion above the pre-stress level.

Large banks (top 25 banks by assets) accounted for 59\% of the C\&I balances since 2005. This share remained stable during the Pandemic onset. Four weeks later, medium and small banks increased their borrowing and pushed the share of large banks permanently below 55\%.

Federal Reserve’s analysis of C\&I draws, see \citet{Glancy2020}, showed that while deposits also increased, many banks had to rely on non‐deposit borrowing to fund the unprecedented credit line utilization. Without incorporating a credit-sensitive spread such as AXI, banks could suffer significant losses during stress periods and banks at risk.

Based on FDIC analysis of the initial COVID‐19 episode, large domestic banks saw companies draw down on their existing revolving credit lines. During a four‐week span at the onset of the pandemic, revolving (C\&I) loan drawdowns increased by nearly \$350 billion. This surge reflected firms rapidly increasing utilization of their prearranged lines of credit to cover funding shortfalls amid the economic shock.

\subsubsection{SVB collapse}
In March 2023, the U.S. experienced the second and third largest bank failures in its history. On March 9, the 16th largest bank in the U.S. with \$209 billion in assets, Silicon Valley Bank (``SVB''), lost 25 percent of deposits and was closed the following day, see \cite{sablik2024}. On March 10, the 29th largest bank in the U.S. with \$110 billion in assets, Signature Bank, lost 20 percent of its deposits, and it was closed on March 12.\footnote{For comparison, it took 16 days for the largest U.S. bank failure, Washington Mutual, to lose 10 percent of its deposits.}

In contrast to the Pandemic onset, major media started speculating about SVB's vulnerabilities and highlighting concerns about its liquidity position only on March 8, 2023. Hence, there is no period of ``build-up'' in AXI when market participants respond in anticipation of future actions.

Unlike during the Pandemic onset, banks responded mainly by increasing their cash positions financed through the Federal Reserve's emergency lending facilities.\footnote{Bank borrowings in Federal Reserve's table H.8 include unsecured overnight borrowing in the federal funds market, reverse repurchase agreements, and term borrowing from commercial banks.} Borrowings jumped \$504 billion from \$1,980 to \$2,484 billion during the week of March 8. About 50\% of this increase was funded through the Federal Home Loan Banks (``FHLB''): the outstanding balances at FHLB increased by \$247 billion and remained elevated until July 2023. The FHLB rate increased around the collapse, and, like during the Pandemic onset, it remained closely tethered to the Federal Funds rate. The reason is they are specific to secured borrowings by member financial institutions, making them less reflective of general funding conditions.

While banks were able to secure the Federal Reserve and FHLB funding, it may not always be the case. First, regulators could restrict funding to banks with rapidly deteriorating credit quality or capital positions. Second, FHLB advances may be further restricted if banks have already pledged their eligible mortgage collateral, if collateral value or quality suffered or market panicked. Finally, while unlikely, FHLB funding capacity may be reached. Indeed, on the first two business days after the SVB collapse, banks probed the fixed income markets for long-term funding, as the AXI long-term dollar transaction volume increased by a factor of 2.8 and 4.8 relative to the respective previous 21-business day averages. The short-term transaction volume was, respectively, 20\% and 30\% lower.

Importantly, the top 25 banks accounted for \$241 billion of the borrowings increase -- less than their size suggests -- as they experienced \$110 billion inflow of deposits in contrast to a \emph{\$236 billion deposit outflow from the non-top 25 banks}.

%\begin{figure}[!htb]
%	\centering
%	\includegraphics{fig-borrowings.eps}
%	\caption{Borrowings}
%	\label{fig-borrowings}
%\end{figure}

%\begin{figure}[!htb]
%	\centering
%	\includegraphics{fig-fhlb_rate.pdf}
%	\caption{FHLB daily reset rate}
%	\label{fig-fhlb_rate}
%\end{figure}

\section{Economic value of AXI}
\label{sec-axi-value}

\subsection{Historical stress scenarios}
This section compares two loan and credit line pricing schemes. The first is the current status quo of basing credit lines to the SOFR 30-day compounded average. The second is the hypothetical loan based on SOFR 30-day compounded average plus AXI. The third option is a 1-month LIBOR loan.

The SOFR+AXI loan includes a fixed spread of 1\%.

\begin{table}[!h]
	\centering
	\begin{tabular}[]{lc}
		Credit line & \$1 million \\
		Rate        & SOFR 30-day compounded average + AXI \\
		Spread      & 1.00\% \\
		Day count   & 360 \\
		Amortized   & No \\
	\end{tabular}
\end{table}

\begin{figure}[!htb]
	\centering
	\includegraphics{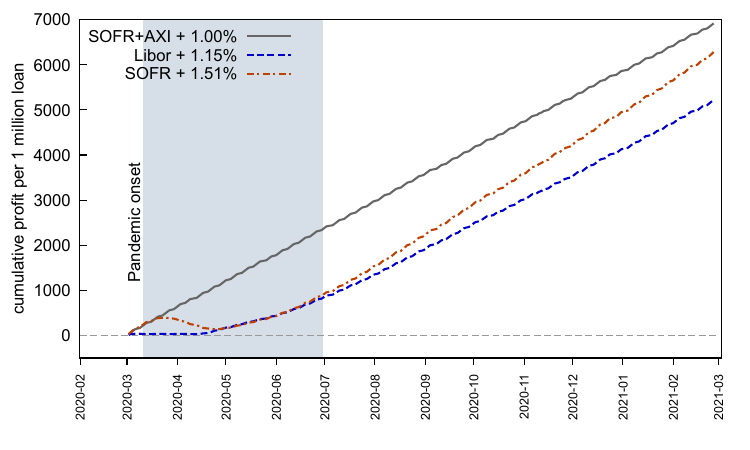}
	\caption{Cumulative credit line profit SOFR+AXI}
	\label{fig-cost-pandemic}
\end{figure}

The spread for LIBOR and SOFR are computed so that the expected income for each pricing scheme is the same:
\begin{align*}
	\textit{average}(\text{SOFR+AXI})+1\%
		&= \textit{average}(\text{SOFR})+1.514\% \\
		&= \textit{average}(\text{LIBOR})+1.148\%.
\end{align*}
Thus, each pricing scheme is expected to earn 1\% annually if SOFR+AXI is the true cost of bank funds. Realized profit will depend on the actual path of rates.
\begin{equation*}
	\textit{Cumulative profit}_{t+s} =
		\sum_{j=1}^{s} \frac{\text{Loan rate}_{t+j}-\text{SOFR+AXI}_{t+j}}{360} \cdot \textit{loan amount}.
\end{equation*}
Assuming interest is paid daily, Figure \ref{fig-cost-pandemic} plots the cumulative profit of a creditor loaning \$1 million on March 1, 2020, i.e., during the Pandemic Onset. SOFR+AXI loan accumulates \$10,000/360 of profit every day. In contrast, LIBOR loan does not generate income until the beginning of May when the funding cost declines sufficiently. SOFR loan would perform comparably to the SOFR+AXI loan before the stress begins but most of the profit would be erased before the beginning of May. Alternatively, banks need to charge higher rates to compensate for their own funding risk.

\begin{table}[!h]
	\centering
    \small
	\begin{tabular}[]{lccc}
		Stress period  & 1 month & \textbf{3 months} & 12 months \\
	\hline
		Pandemic onset, SOFR+AXI vs. SOFR  & 2.5 bp & \textbf{13.4 bp} &  6.4 bp\\
		Pandemic onset, SOFR+AXI vs. LIBOR & 5.7 bp & \textbf{13.1 bp} & 16.1 bp\\
	\hline
		  SVB collapse, SOFR+AXI vs. SOFR  & 5.1 bp & \textbf{14.8 bp} & 14.1 bp\\
		  SVB collapse, SOFR+AXI vs. LIBOR & 4.3 bp & \textbf{12.8 bp} &  9.2 bp\\
	\end{tabular}
	\caption{Additional profits from using credit-sensitive rates during stress periods}
	\label{tbl-credit_sens_profits}
\end{table}

Table~\ref{tbl-credit_sens_profits} reports the additional profit earned from using the credit-sensitive rate SOFR+AXI during the two stress periods after 1, 3, and 12 month period. Focusing on 3 months, the additional profit ranges between 13.1 and 14.8 basis points. These numbers are economically significant: when applied to the additional C\&I balances drawn during the Pandemic Onset, the additional interest earned by the banking sector over three months would be $ \$370\ \text{billion} \cdot 0.134\% = \$0.496 $ billion. For comparison, it is 2.7\% of the total bank profit in Q1 2020, which was \$18.5 billion, see \citet{FDIC2020Q1}. When applied to the total unused commitments, the additional interest earned over three months would be \$4,829 billion $ \cdot $ 0.134\% = \$6.471 billion, approximately 34\% of the total banking profit in the same quarter.

%\begin{align*}
%	\textit{profit}(\text{SOFR+AXI})_t &= 1\% \\
%	\textit{profit}(\text{SOFR})_t &= 1.52\% - \text{AXI}_t \\
%	\textit{profit}(\text{Libor})_t &= 1.33\% + \text{Libor}_t - (\text{SOFR+AXI})_t \\
%\end{align*}

%\begin{figure}[!htb]
%	\centering
%	\includegraphics{fig-cost-pandemic.pdf}
%	\caption{Additional interest earned when indexing loans to SOFR+AXI}
%	\label{fig-cost-pandemic}
%\end{figure}

\subsection{Risk-adjusted return}
The two loan/credit line indexing schemes can be compared using the risk adjusted return. Conceptually similar to \citet{Sharpe1966} ratio, it captures expected return per unit of risk and is defined as:
\begin{equation*}
	\textit{risk-adjusted return} = \frac{\textit{expected excess return}}{\textit{return volatility}}.
\end{equation*}
% (R1-Rf)/S1 = (R2-Rf)/S2 -> R2 = Rf+(R1-Rf)S2/S1

Banks that offer loans indexed to SOFR + AXI assume less funding risk than banks that index loans solely to SOFR. Consequently, they can offer lower loan rates while maintaining the same risk-adjusted return. The magnitude of this rate discount reflects the value of the funding protection embedded in AXI and can be quantified as follows.\footnote{This is consistent with the arguments in \citet{Cooperman2025} and \citet{Ghamami2023} that loans based on risk-free rates would be more expensive on average.}

Suppose a bank offers loans with a rate structure given by:
\begin{equation*}
	\textit{Loan rate}_t = R_t + s + c \cdot \text{AXI}_t.
\end{equation*}
where $ R_t $ is a risk-free rate benchmark, e.g., SOFR 30-day compounded average, $ s $ is a fixed spread, and $ c $ is credit sensitivity bounded between 0 and 1. Importantly, $ \text{AXI}_t $ is \emph{expected rate at which an average bank borrows.}

Cost of funds to bank $ i $ in period $ t $ is:
\begin{equation*}
	\textit{Cost of funds}_{ti} = R_t + \text{AXI}_t + \Delta_{ti}.
\end{equation*}
where $\Delta_{ti}$ captures bank-specific deviations from the weighted average funding rate, i.e., AXI.\footnote{The definition of AXI implies $\sum_i \text{maturity}_{ti} \cdot \text{amount}_{ti} \Delta_{ti} \approx 0 $, as confirmed in Table~\ref{tbl-params_rar}.} For example, banks relying on short-term funding typically borrow at rates below AXI, while banks reliant on longer-term funding face higher borrowing costs relative to AXI.

Loan profitability is the relevant excess return. It is risky, since the realized cost of funds differs from AXI and varies across banks and over time:
\begin{equation*}
	\textit{Profitability}_{ti} = s + (c-1) \cdot \text{AXI}_t - \Delta_{ti}.
\end{equation*}
Expected risk-adjusted return is:
\begin{equation}
	\text{Expected RAR} = \frac{s + (c-1) \cdot \overline{\text{AXI}}}{\sqrt{(c-1)^2 \cdot \sigma^2(\text{AXI}) + \sigma^2(\Delta)}}.
	\label{eq-rar}
\end{equation}
$ \sigma^2(\text{AXI}) $ and $ \sigma^2(\Delta) $ denote variance of AXI and bank specific deviation, respectively, and $ \overline{\text{AXI}} $ denotes the mean AXI. Equation~\eqref{eq-rar} assumes that deviations from AXI and $ \Delta $ are uncorrelated: $ cor(\text{AXI}_t,\Delta_{ti})=0 $. Practically, it rules out a possibility that the distribution of deviations widens when AXI increases.

As a bank increases credit sensitivity, expected profitability increases and its volatility decreases. The maximum risk-adjusted return $ \text{RAR}_1 = s/\sigma(\Delta) $ is achieved at full credit sensitivity: $ c=1 $. The risk-adjusted return is $ \text{RAR}_0 = [s - \overline{\text{AXI}}]/\sqrt{\sigma^2(\text{AXI}) + \sigma^2(\Delta)} $, without the credit-sensitive component, $ c=0 $. With credit-sensitive rate structure, banks can lower $ s $ without sacrificing risk-adjusted return.

The spread for a credit-sensitive-rate loan $ s' $ that maintains the same risk-adjusted return as the SOFR-only loan with spread $ s $ is:
\begin{equation}
	s'
	= (1-c)\cdot \overline{\text{AXI}} + [s - \overline{\text{AXI}}] \frac{\sqrt{(c-1)^2 \cdot \sigma^2(\text{AXI}) + \sigma^2(\Delta)}}{\sqrt{\sigma^2(\text{AXI}) + \sigma^2(\Delta)}}.
	\label{eq-spread_SOFR+AXI}
	\normalsize
\end{equation}
The term $ (1-c)\cdot \overline{\text{AXI}} $ is the average unhedged portion of the funding cost, and $ s-\overline{\text{AXI}} $ is the portion of AXI offset by the fixed spread of the SOFR-only loan. The ratio, which is always between 0 and 1, adjusts for the relative volatility of loan profitability under SOFR-only and credit-sensitive loan pricing. In the example calibrated to June 2016-April 2025, the bank can cut its contractual spread nearly two-thirds without lowering risk-adjusted returns.

\begin{table}[!htb]
	\centering
	\begin{tabular}{lcccc}
		parameter & $ \overline{\text{AXI}} $ & $ \sigma(\text{AXI}) $ & $ \bar{\Delta} $ & $ \sigma(\Delta) $ \\
		\hline
		%RETIRED value     & 0.5165 & 0.2999 &    n.a. & 0.1350 \\
		value     & 0.5141\% & 0.2987\% & -0.0020\% & 0.3156\% \\
	\end{tabular}
	\caption{Assumed parameters for potential spread discount}
	\label{tbl-params_rar}
\end{table}

\begin{figure}[!htb]
	\centering
	\includegraphics{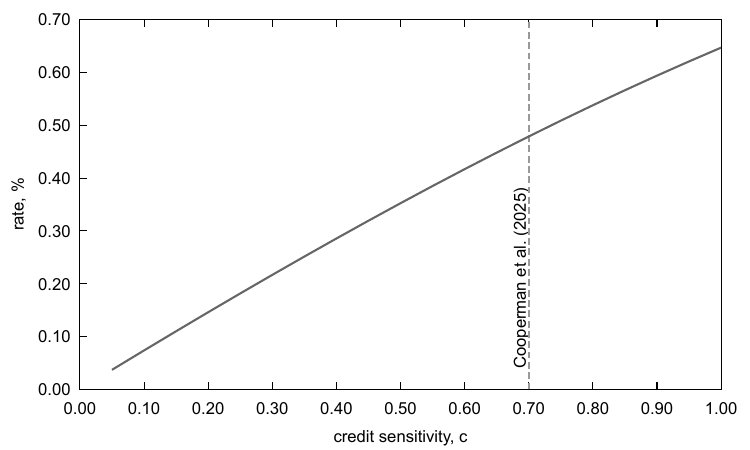}
	\caption{Potential spread reduction when switching from SOFR to SOFR+AXI pricing while maintaining risk-adjusted return}
	\label{fig-spread_reduction}
\end{figure}

Assuming the initial fixed spread under SOFR-only pricing is $ s=1\% $, Figure \ref{fig-spread_reduction} plots the rate discount $ s-s' $ based on equation \eqref{eq-spread_SOFR+AXI} for different levels of credit sensitivity. Table~\ref{tbl-params_rar} reports parameter assumptions. $ \sigma(\Delta) $ is estimated assuming that $ \Delta $ fluctuates between two values: long-term and short-term spread with probabilities implied by the respective dollar volumes, see Appendix \ref{sec-sigma_Delta}. \citet{Cooperman2025} estimate the optimal credit sensitivity at $c=0.70$, at which level a bank could reduce spreads by approximately 48 basis points, offering a 52 basis point spread instead of 100 basis points. This represents an economically meaningful reduction in loan borrowing cost, or alternatively potential improvement in risk-adjusted return.

The preceding analysis abstracts from potential effects on loan demand. In practice, shifting interest rate risk to borrowers through a credit-sensitive rate structure will reduce demand for loans. By lowering the fixed spread, banks can offset some of the cost borne by borrowers and preserve credit demand. For example, using the price elasticity of 25 assumed in \citet{Cooperman2022}, the impact of reducing the fixed spread by 48 basis points is to increase demand by 12.0\%.

\section{Conclusions}
In conclusion, the analysis presented underscores the significant economic benefits of adopting a credit-sensitive rate structure such as SOFR+AXI, particularly during periods of financial stress. By incorporating AXI, banks would align loan rates with their true marginal funding costs. Historical stress scenarios, including the COVID-19 pandemic onset and the Silicon Valley Bank collapse, clearly illustrate that reliance solely on risk-free rates like SOFR can lead to considerable funding mismatches and profitability erosion.

Incorporating AXI into loan pricing could significantly reduce banks' exposure to sudden spikes in funding costs. Banks adopting credit-sensitive rates can offer meaningful discounts to borrowers, potentially reducing borrowing costs by up to 65 basis points and benefitting the broader economy while maintaining risk-adjusted returns.

Finally, adherence to IOSCO principles, further reinforces AXI's viability as a foundational benchmark in a post-LIBOR landscape. Given these compelling advantages, banking institutions, market participants, and regulators should consider adopting AXI, recognizing its role in managing funding risk and promoting resilience in financial markets.

\bibliography{axi-empirical-study}

% LENDING THROUGH THE PANDEMIC AND THE RECOVERY

%ALTERNATIVE REFERENCE RATES COMMITTEE, May 21, 2021, ARRC Releases Update on its RFP Process for Selecting a Forward-Looking SOFR Term Rate Administrator, \url{https://www.newyorkfed.org/medialibrary/Microsites/arrc/files/2021/20210521-ARRC-Press-Release-Term-Rate-RFP.pdf}
%
%LIBOR Legacy Playbook link, \url{https://www.newyorkfed.org/medialibrary/Microsites/arrc/files/2022/LIBOR_Legacy_Playbook.pdf}
%
%Progress Report: The Transition from U.S. Dollar LIBOR, The Alternative Reference Rates Committee, March 2021 (Updated as of March 31, 2021), \url{https://www.newyorkfed.org/medialibrary/Microsites/arrc/files/2021/USD-LIBOR-transition-progress-report-mar-21.pdf}
%
%FDIC, (2020). ``Consumer Lending Through The Pandemic and The Recovery.'' Available at \url{https://www.fdic.gov/analysis/quarterly-banking-profile/fdic-quarterly/2022-vol16-1/article1.pdf}
%
%Federal Reserve, (2020). ``How Did Banks Fund C\&I Drawdowns?'' Availabel at \url{https://www.federalreserve.gov/econres/notes/feds-notes/how-did-banks-fund-c-and-i-drawdowns-at-the-onset-of-the-covid-19-crisis-20200731.html}

\appendix
\small

\begin{figure}[!htb]
	\includegraphics[]{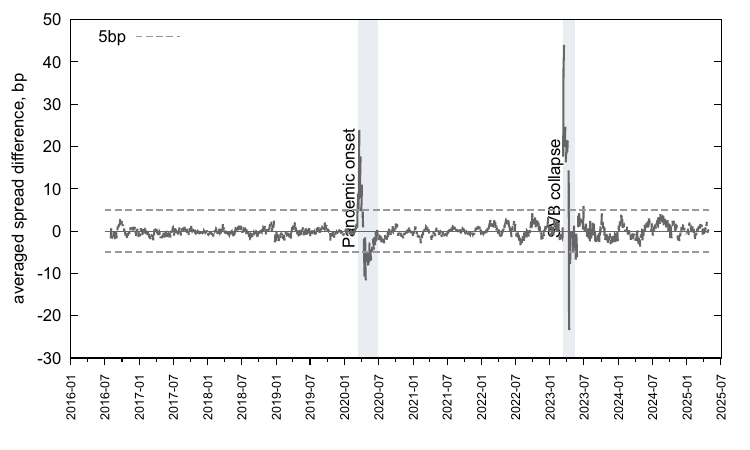}
	\caption{AXI: Difference between 21-business day simple average and 30-day compound average}
	\label{fig-axi-avgdiff}
\end{figure}

\section{Credit spread variability facing individual banks}
\label{sec-sigma_Delta}
AXI is the average spread that can be used by banks to price loans, while each individual bank's funding cost differs by $ \Delta_{ti} $, where $ i $ and $ t $ denote a bank and transaction day, respectively. Without the full distribution of $ \Delta_{ti} $, this analysis assumes that $ \Delta_{ti} $ takes only two values: $ \text{LT spread}_t-\text{AXI}_{t} $ or $ \text{ST spread}_t-\text{AXI}_{t} $. This corresponds to the thought experiment in which a bank finds out whether it needs short- or long-term funding in the beginning of each day.

The probabilities of the two realizations equal the maturity-weighted shares of the long-term and short-term transaction volume, respectively. This specification views a $ N $-day loan as $ N $ one-day loans of the same amount.

The estimate of $ \sigma(\Delta_{ti}) $ is the average of the daily (Bernoulli) volatility estimate:
\begin{equation*}
	\sigma(\Delta_{ti}) = \frac{1}{T} \sum_{t=1}^T \sqrt{\text{LT weight}_t\cdot \text{ST weight}_t} \cdot |\text{LT spread}_t-\text{ST spread}_t|.
\end{equation*}
Using unweigted dollar volumes to estimate probabilities would result in a lower $ \sigma(\Delta_{ti}) $ estimate and a larger potential spread reduction. That is, the assumptions above are conservative.

\section{AXI's monthly and daily correlations}
\label{s-correlations-fin-extended}

Table~\ref{tbl-correlations-fin-extended} reports monthly correlations between changes in AXI and changes in selected indicators. The strongest co-movements are with the two OAS measures (0.714 and 0.670), LIBOR–OIS (0.617), the TED spread (0.611), and the NFCI (0.622). For NFCI, the contemporaneous correlation (0.595) is statistically indistinguishable from its peak (0.622) when AXI lags NFCI by one month, implying that at the monthly frequency AXI and NFCI move largely contemporaneously. This is consistent with section~\ref{sec-axi-drivers}, which shows that AXI lags the same indicators by one week.

Table~\ref{tbl-correlations-fin-extended-daily} reports daily correlations. The two OAS indices the strongest correlations with AXI (0.280 and 0.274) at a one-day lag. All series but BFRN OAS (the correlations are statistically indistinguishable across lags) attain their strongest correlation with AXI when the latter lags by five trading days (not shown), consistent with AXI’s one-week lag documented in section~\ref{sec-axi-drivers}.

\begin{table}[!htb]
	\centering
	\small
	\begin{tabular}{lcr|rrr}
		variable & stat. & current & lag 1 & lag 2 & lag 3 \\
		\hline
		Markit NA credit index
			& correl. & \color{blue}{\textbf{0.380}} & \textbf{0.323} & \textbf{0.266} & \textbf{0.209} \\
			& p-value & 0.000 & 0.000 & 0.000 & 0.000 \\
		Bloomberg FRN index, OAS
			& correl. & \color{blue}{\textbf{0.714}} & \textbf{0.512} & \textbf{0.310} & \textbf{0.107}\\
			& p-value & 0.000 & 0.000 & 0.000 & 0.000\\
		Bloomberg FRN index
			& correl. & \color{blue}{\textbf{-0.623}} & \textbf{-0.445} & \textbf{-0.267} & \textbf{-0.088}\\
			& p-value & 0.000 & 0.000 & 0.000 & 0.058 \\
		Bloomberg US corporate index, OAS
			& correl. & \color{blue}{\textbf{0.670}} & \textbf{0.498} & \textbf{0.327} & \textbf{0.155}\\
			& p-value & 0.000 & 0.000 & 0.000 & 0.001 \\
		Bloomberg US corporate index
			& correl. & \color{blue}{\textbf{-0.276}} & \textbf{-0.233} & \textbf{-0.190} & \textbf{-0.147}\\
			& p-value & 0.000 & 0.000 & 0.000 & 0.002 \\
		National financial conditions index
		    & correl. & \textbf{0.595} & \color{blue}{\textbf{0.622}} & \textbf{0.534} & \textbf{0.368}\\
		    & p-value & 0.000 & 0.000 & 0.000 & 0.000 \\
		Credit market distress index
		    & correl. & \color{blue}{\textbf{0.309}} & \textbf{0.173} & \textbf{0.150} & \textbf{-0.025} \\
		    & p-value & 0.000 & 0.000 & 0.001 & 0.593 \\
        LIBOR-OIS spread
		    & correl. & \color{blue}{\textbf{0.617}} & \textbf{0.443} & \textbf{0.269} & \textbf{0.095} \\
		    & p-value & 0.000 & 0.000 & 0.000 & 0.042 \\
        TED spread
		    & correl. & \color{blue}{\textbf{0.611}} & \textbf{0.449} & \textbf{0.288} & \textbf{0.126} \\
		    & p-value & 0.000 & 0.000 & 0.000 & 0.007 \\
		\hline
		SPDR financial sector ETF
			& correl. & \color{blue}{\textbf{-0.426}} & \textbf{-0.344} & \textbf{-0.262} & \textbf{-0.180}\\
			& p-value & 0.000 & 0.000 & 0.000 & 0.000 \\
		\hline
		Economic policy uncertainty
		    & correl. & \color{blue}{\textbf{0.244}} & \textbf{0.184} & \textbf{0.123} & 0.062 \\
		    & p-val. & 0.000 & 0.000 & 0.009 & 0.184 \\
		Volatility index
			& correl. & \textbf{0.103} & \color{blue}{\textbf{0.146}} & \textbf{0.188} & \textbf{0.230}\\
			& p-value & 0.026 & 0.002 & 0.005 & 0.000 \\
	\end{tabular}

 	\caption{Monthly correlations of AXI with selected financial indicators}
	\label{tbl-correlations-fin-extended}
\end{table}

\begin{table}[!htb]
	\centering
	\small
	\begin{tabular}{lcrr|rr}
		variable & stat. & current & lag 1 & lag 2 & lag 3 \\
		\hline
		Markit NA credit index
			& correl. & -0.015 & 0.018 & \textbf{0.076} & \color{blue}{\textbf{0.094}}\\
			& p-value & 0.489 & 0.403 & 0.000 & 0.000 \\
		Bloomberg FRN index OAS
			& correl. & \textbf{0.275} & \color{blue}{\textbf{0.280}} & \textbf{0.275} & \textbf{0.275}\\
			& p-value & 0.000 & 0.000 & 0.000 & 0.000\\
		Bloomberg FRN index
			& correl. & \textbf{-0.298} & \textbf{-0.316} & \textbf{-0.329} & \color{blue}{\textbf{-0.342}}\\
			& p-value & 0.000 & 0.000 & 0.000 & 0.000 \\
		Bloomberg US corporate index OAS
			& correl. & 0.196 & \textbf{0.274} & \textbf{0.316} & \color{blue}{\textbf{0.326}}\\
			& p-value & 0.000 & 0.000 & 0.000 & 0.000 \\
		Bloomberg US corporate index
			& correl. & -0.036 & \textbf{-0.038} & \textbf{-0.064} & \color{blue}{\textbf{-0.084}}\\
			& p-value & 0.090 & 0.075 & 0.003 & 0.000 \\
        LIBOR-OIS spread
		    & correl. & \textbf{0.134} & \textbf{0.167} & \textbf{.185} & \color{blue}{\textbf{0.210}}\\
		    & p-value & 0.000 & 0.000 & 0.000 & 0.000 \\
        TED spread
		    & correl. & \textbf{0.121} & \textbf{0.132} & \textbf{0.125} & \color{blue}{\textbf{0.154}}\\
		    & p-value & 0.000 & 0.000 & 0.000 & 0.000 \\
		\hline
		SPDR financial sector ETF
			& correl. & -0.033 & \textbf{-0.058} & \textbf{-0.099} & \color{blue}{\textbf{-0.111}}\\
			& p-value & 0.126 & 0.006 & 0.000 & 0.000 \\
		\hline
		Economic policy uncertainty
		    & correl. & 0.002 & 0.022 & 0.003 & 0.020\\
		    & p-val. & 0.933 & 0.313 & 0.881 & 0.346\\
		Volatility index
			& correl. & -0.006 & 0.012 & 0.033 & 0.035\\
			& p-value &  0.766 & 0.558 & 0.122 & 0.103\\
	\end{tabular}

	\caption{Daily correlations of AXI with selected financial indicators}
	\label{tbl-correlations-fin-extended-daily}
\end{table}

\newpage
Table \ref{tbl-axi-drivers} lists the data sources for the considered indexes. Several indexes not listed below were considered but their results excluded for lack of meaningful relationships with AXI, e.g., recession probability from the Federal Reserve Bank of New York.

\begin{table}[!h]
	\centering
	\small
	\begin{tabular}{lcr}
		Indicator & Source & Code\\
		\hline
		Markit NA credit index & Bloomberg & CDXNAIG \\
		Bloomberg FRN index & Bloomberg & BFU5TRUU \\
%		Bloomberg FRN index & Bloomberg & BFRNTRUU \\
%		Bloomberg US credit index & Bloomberg & LBUSTRUU \\
		Bloomberg US corporate index & Bloomberg & LDC5TRUU \\
		Credit market distress index & FRB of New York & CDMIIG \\
		\hline
		National financial conditions index & FRB of Chicago & NFCI\\
		Economic policy uncertainty index & EPU webpage & EPU \\
		Volatility index & Bloomberg & VIX \\
		\hline
		SPDR financial sector ETF index & Bloomberg & XLF \\
		\hline
        LIBOR-OIS spread & FRED & LIBOROIS\\
		TED spread & FRED & TEDSPREAD \\
	\end{tabular}

	\caption{List of potential AXI drivers}
	\label{tbl-axi-drivers}
\end{table}

\end{document}